\documentclass[aps,pra,preprint,superscriptaddress]{revtex4-1}

\usepackage{amsmath,amssymb,graphicx}
\usepackage[english]{babel}
\usepackage{graphicx}

\begin{document}

\title{Exceptional solutions in two-mode quantum Rabi models}

\author{S. A. Chilingaryan}
\affiliation{Departamento de F\'{i}sica, Universidade Federal de Minas Gerais, Caixa Postal 702, 30123-970, Belo Horizonte, MG, Brazil}

\author{B. M. Rodr\'{\i}guez-Lara}
\email{bmlara@inaoep.mx}
\affiliation{Instituto Nacional de Astrof\'{\i}sica, \'Optica y Electr\'onica, Calle Luis Enrique Erro No. 1, Sta. Ma. Tonantzintla, Pue. CP 72840, M\'exico}

\begin{abstract}
We study two models describing the interaction of a two-level system with two quantum field modes. 
The first one is equivalent to a dissipative two-state system with just two boson fields in the absence of tunneling. 
The second describes two orthogonal fields interacting with the corresponding orthogonal dipoles of a two-level system.
We show that both models present a partial two-mode $SU(2)$ symmetry and that they can be solved in the exceptional case of resonant fields. 
We study their ground state configurations, that is, we find the quantum precursors of the corresponding semi-classical phase transitions, as well as their whole spectra to infer their integrability. 
We show that the first model in the exceptional case is isomorphic with the quantum Rabi model and allows just two ground state configurations, vacuum and non-vacuum.
The second model allows four ground state configurations, one vacuum, two non-vacuum single mode and one non-vacuum dual mode, and give analytic and numerical pointers that may suggest its integrability.
We also show that in the single excitation subspace these models can serve as a fast $SU(2)$ beam splitter even in the ultra-strong coupling regime.
\end{abstract}


\maketitle
\section{Introduction} \label{sec:S1}

The so-called quantum Rabi model \cite{Rabi1936p324}, 
\begin{eqnarray}
\hat{H}_{R} = \frac{\omega_{0}}{2} \hat{\sigma}_{z} + \omega \hat{a}^{\dagger} \hat{a} + g \left( \hat{a}^{\dagger} + \hat{a} \right) \hat{\sigma}_{x},
\end{eqnarray}
modeling the interaction of a two-level system, described by the transition frequency $\omega_{0}$ and the Pauli operators $\hat{\sigma}_{j}$ with $j=x,y,z$, with a boson field, described by the field frequency $\omega$ and the annihilation (creation) operators $\hat{a}$ ($\hat{a}^{\dagger}$) can be seen as a single-field version of the dissipative two-state system  \cite{Leggett1987p1} in the absence of tunneling, $\Delta=0$, 
\begin{eqnarray}
\hat{H}_{L} = \frac{\omega_{0}}{2} \hat{\sigma}_{z} - \frac{\Delta}{2} \hat{\sigma}_{x} + \sum_{j} \omega_{j} \hat{a}_{j}^{\dagger} \hat{a}_{j} + \sum_{j} g_{j} \left( \hat{a}_{j}^{\dagger} + \hat{a}_{j} \right) \hat{\sigma}_{x}.
\end{eqnarray}
The dissipative two-state model is characterized by a spectral function, $J(\omega) = \pi \sum_{j} g_{j}^{2} \delta(\omega_{j} - \omega)$ and is solvable, for example, for sub-Ohmic, Ohmic and super-Ohmic spectral functions, $J(\omega) \propto \omega^{s}$ with $s<1$, $s=1$ and $s>1$, in that order \cite{Leggett1987p1}.
On the other hand, the solvability and integrability of the quantum Rabi model has been recently discussed for any given parameter set \cite{Braak2011p100401,Moroz2012p60010,Moroz2013p319,Braak2013p175301,Braak2013p23,Moroz2014p252}.
An equivalent approach has been used to explore the integrability and exceptional solutions of the two-qubit quantum Rabi model \cite{Peng2012p365302,Chilingaryan2013p335301,Moroz2014p960,RodriguezLara2014p135306,Peng2014p265303}.

Here, we are interested in exploring the exceptional solutions of models describing a single qubit coupled to just two boson fields. 
Our motivation is twofold. 
First, the use of Bargmann approach may render the system integrable for exceptional parameter sets in a way equivalent to that found in the two-qubit single-field case. 
Second, circuit quantum electrodynamics (circuit-QED) may provide a direct testing ground for such a model from weak to ultrastrong couplings \cite{Baust2014}, 
\begin{eqnarray}
\hat{H}_{1} = \frac{\omega_{0}}{2} \hat{\sigma}_{z} + \sum_{j=1}^{2} \omega_{j} \hat{a}_{j}^{\dagger} \hat{a}_{j} + \sum_{j=1}^{2} g_{j} \left( \hat{a}_{j}^{\dagger} + \hat{a}_{j} \right) \hat{\sigma}_{x}
\end{eqnarray} 
and cavity-QED may provide an equivalent model by  Raman adiabatic driving of a four-level atom coupled to two cavity electromagnetic field modes \cite{Fan2014p023812},
\begin{eqnarray}
\hat{H}_{2} = \frac{\omega_{0}}{2} \hat{\sigma}_{z} + \sum_{j=1}^{2} \omega_{j} \hat{a}_{j}^{\dagger} \hat{a}_{j} +  g_{1} \left( \hat{a}_{1}^{\dagger} + \hat{a}_{1} \right) \hat{\sigma}_{x} + i g_{2} \left( \hat{a}_{2}^{\dagger} - \hat{a}_{2} \right) \hat{\sigma}_{y}.
\end{eqnarray} 
Note that both models conserve parity, $\hat{\Pi}= e^{- i \pi \hat{N}}$ defined in terms of the total number of excitation $\hat{N} = \hat{\sigma}_{z}/2 + \hat{a}_{1}^{\dagger} \hat{a}_{1} + \hat{a}_{2}^{\dagger} \hat{a}_{2} + 1/2$.

This manuscript is structured as follows. 
First, we will study possible candidates for symmetries on these models and regimes where they are equivalent to well known models.
Here, we will introduce the exceptional case provided by resonant fields, where Hamiltonian $\hat{H}_{1}$ reduces to the standard quantum Rabi model and Hamiltonian $\hat{H}_{2}$ is invariant to a $SU(2) \otimes SU(2)$ transformation for identical couplings.
Then, we will focus on this exceptional case of resonant fields to study the ground state structure of both models. 
While Hamiltonian $\hat{H}_{1}$ shows a simple ground state configuration that includes just a vacuum and a non-vacuum ground states, Hamiltonian $\hat{H}_{2}$ shows a more interesting ground state configuration landscape with four possible configurations, one vacuum, two non-vacuum single mode, and one non-vacuum dual mode.
Next, we will discuss the integrability of Hamiltonian $\hat{H}_{1}$ due to the isomorphism with the quantum Rabi model and give analytic and numeric arguments that point in the same direction for Hamiltonian $\hat{H}_{2}$.
Finally, we will demonstrate that the partial $SU(2)$ symmetry, shown by both models in the exceptional case of resonant fields, allows us to construct closed form evolution operators in the weak coupling regime. 
We will use these evolution operators to show that these models may be used as $SU(2)$ beam splitters in the single excitation subspace and, finally, we will present a couple of exact numeric time evolution that show this beam splitter effect survives even in the ultra-strong coupling regime for short evolution times.

\section{Symmetries and equivalence with other models.}

First, we want to bring forward that both Hamiltonians $\hat{H}_{1}$ and $\hat{H}_{2}$ are invariant to full rotations, $\theta= 2 n \pi$ with $n=0,1,2,\ldots$, under the unitary transformation, 
\begin{eqnarray}
\hat{U}(\theta) = e^{i \theta \left( \hat{a}_{1}^{\dagger} \hat{a}_{2} + \hat{a}_{1} \hat{a}_{2}^{\dagger} - \hat{\sigma}_{z} /2 \right)},
\end{eqnarray}
in other words, 
\begin{eqnarray}
\hat{U}(2 n \pi) \hat{H}_{j} \hat{U}^{\dagger}(2 n \pi) = \hat{H}_{j}.
\end{eqnarray}
The field part of this transformation is related to Schwinger two-mode representation of SU(2)  \cite{Biedenharn1965}, $\hat{J}_{+} = \hat{a}_{1}^{\dagger} \hat{a}_{2}$, $\hat{J}_{-} = \hat{a}_{1} \hat{a}_{2}^{\dagger}$, $\hat{J}_{0} = (\hat{a}_{1}^{\dagger}\hat{a}_{1} - \hat{a}_{2}^{\dagger} \hat{a}_{2})/2$.
This does not provide us with any information but note that in the case of identical qubit-field couplings, $g_{1}=g_{2}$, Hamiltonian $\hat{H}_{2}$ is invariant to any given rotation parameter \cite{Fan2014p023812}, 
\begin{eqnarray}
\hat{U}(\theta) \hat{H}_{2} \hat{U}^{\dagger}(\theta) = \hat{H}_{2}, \quad g_{1} = g_{2}.
\end{eqnarray}

On the other hand, it is well known that there exists an exact unitary transformation that maps the dissipative two-level model, $\hat{H}_{L}$, into a linear nearest neighbor chain of coupled bosonic modes where just the first one of them is coupled to the qubit \cite{Chin2010p092109}. 
It is not surprising that such transformation in the finite case is related to the  SU(2) unitary displacement operator,
\begin{eqnarray}
\hat{D}(\xi) = e^{\xi \left( \hat{a}_{1}^{\dagger} \hat{a}_{2} - \hat{a}_{1} \hat{a}_{2}^{\dagger} \right)}, \quad \tan \xi = \frac{g_{2}}{g_{1}}.
\end{eqnarray}
This two-mode displacement yields an effective model where the qubit couples to only the first boson field in the usual quantum Rabi model form, and the first and second boson field couple between them with a beam splitter form,
\begin{eqnarray}
\hat{H}_{1D} &=& \hat{D}(\xi) \hat{H}_{1} \hat{D}^{\dagger}(\xi), \\
&=& \frac{\omega_{0}}{2} \hat{\sigma}_{z} + \sum_{j=1}^{2} \Omega_{j} \hat{a}_{j}^{\dagger} \hat{a}_{j} + \lambda \left( \hat{a}^{\dagger}_{1} \hat{a}_{2}  + \hat{a}_{1}  \hat{a}^{\dagger}_{2}  \right)  + g \left( \hat{a}_{1}^{\dagger} + \hat{a}_{1} \right) \hat{\sigma}_{x}. 
\end{eqnarray}
Here, we have defined effective field frequencies $\Omega_{1} = \left( \omega_{1} g_{1}^{2} + \omega_{2} g_{2}^{2} \right)/g^{2}$ and $\Omega_{2} = \left( \omega_{1} g_{2}^{2} + \omega_{2} g_{1}^{2} \right)/g^{2}$, effective field coupling constant $\lambda = \left( \omega_{2} - \omega_{1} \right) g_{1} g_{2}/ g^{2}$, and effective qubit-field coupling $g = \sqrt{g_{1}^{2} + g_{2}^{2}}$.
Note that choosing $\tan \xi = - g_{1}/ g_{2}$ as two-mode displacement parameter just interchanges the boson field modes.

A set of unitary transformations cannot bring Hamiltonian $\hat{H}_{2}$ into an expression similar to $\hat{H}_{1D}$, thus we are reduced to explore regimes where they may be equivalent. 
We can start by using the same two-mode displacement on $\hat{H}_{2}$ and find, 
\begin{eqnarray}
\hat{H}_{2D}  &=& \hat{D}(\xi) \hat{H}_{2} \hat{D}^{\dagger}(\xi), \\
&=& \frac{\omega_{0}}{2} \hat{\sigma}_{z} + \sum_{j=1}^{2} \Omega_{j} \hat{a}_{j}^{\dagger} \hat{a}_{j} + \lambda \left( \hat{a}^{\dagger}_{1} \hat{a}_{2}  + \hat{a}_{1}  \hat{a}^{\dagger}_{2}  \right)  +  \nonumber \\ 
&& + \left[ g \hat{a}_{1}^{\dagger} + \frac{g_{1}^{2}- g_{2}^{2}}{g} \hat{a}_{1} - \frac{2 g_{1} g_{2}}{g} \hat{a}_{2} \right] \hat{\sigma}_{+} + \left[ \frac{g_{1}^{2}- g_{2}^{2}}{g} \hat{a}_{1}^{\dagger} +  g \hat{a}_{1} - \frac{2 g_{1} g_{2}}{g} \hat{a}_{2}^{\dagger} \right] \hat{\sigma}_{-} ,
\end{eqnarray} 
where the effective frequencies and couplings are the same as in the previous case.
At most, we may obtain a similar form to effective Hamiltonian $\hat{H}_{1D}$ in the somewhat obvious regime $g_{1} \gg 2 g_{2}$ and $g_{2} \gg 0$ where we can approximate, 
\begin{eqnarray}
\hat{H}_{2D}  & \approx & \frac{\omega_{0}}{2} \hat{\sigma}_{z} + \sum_{j=1}^{2} \Omega_{j} \hat{a}_{j}^{\dagger} \hat{a}_{j} + \lambda \left( \hat{a}^{\dagger}_{1} \hat{a}_{2}  + \hat{a}_{1}  \hat{a}^{\dagger}_{2}  \right)  + g_{1} \left( \hat{a}_{1}^{\dagger} + \hat{a}_{1} \right) \hat{\sigma}_{x}.
\end{eqnarray} 
Again, it is possible to interchange the fields via the transformation parameter $\xi$.
Thus, Hamiltonian $\hat{H}_{1}$ will share the properties shown by $\hat{H}_{2}$ in the particular regions $g_{1} \gg 2 g_{2}$ and $g_{2} \gg 2 g_{1}$.
Furthermore, in the case of identical couplings, where $\hat{H}_{2}$ is invariant to rotations $\hat{U}(\theta)$, 
\begin{eqnarray}
\left. \hat{H}_{2D} \right\vert_{g_{1}=g_{2}}  & = & \frac{\omega_{0}}{2} \hat{\sigma}_{z} + \frac{1}{2} \left( \omega_{1} + \omega_{2} \right) \sum_{j=1}^{2} \hat{a}_{j}^{\dagger} \hat{a}_{j} + \left( \omega_{2} - \omega_{1} \right) \left( \hat{a}^{\dagger}_{1} \hat{a}_{2}  + \hat{a}_{1}  \hat{a}^{\dagger}_{2}  \right) + \nonumber \\
&& \sqrt{2} g_{1} \left[ \left( \hat{a}_{1}^{\dagger} - \hat{a}_{2} \right) \hat{\sigma}_{+} + \left( \hat{a}_{1} - \hat{a}_{2}^{\dagger} \right) \hat{\sigma}_{-} \right],
\end{eqnarray} 
This Hamiltonian is equivalent to two resonant fields, one of them interacting under the Jaynes-Cummings dynamics with the qubit and the other under anti-Jaynes-Cummings dynamics \cite{RodriguezLara2005p023811}.
Note that under resonant fields, $\omega_{1} = \omega_{2}$, it also conserves the quantity $\hat{\mathcal{N}} = - \hat{a}_{1}^{\dagger} \hat{a}_{1} + \hat{a}_{2}^{\dagger} \hat{a}_{2} + \hat{\sigma}_{z}/2 + 1/2$ and, thus, can be solved.

\section{Ground state configuration.}

It is well known that the spin-$N/2$ version of Hamiltonian $\hat{H}_{2}$ in the semi-classical limit, $N \gg 1$, allows four types of ground state \cite{Fan2014p023812}. 
These that can be characterized by four order parameters given by the qubit energy difference, $\langle \hat{\sigma}_{z} \rangle$, the two mean photon numbers, $\langle \hat{a}_{j}^{\dagger} \hat{a}_{j} \rangle$, and the mean two mode photon number, $\langle \hat{\chi} \rangle$ with $\hat{\chi} = \left( \hat{a}^{\dagger}_{1} + \hat{a}_{2}^{\dagger} \right)\left( \hat{a}_{1} + \hat{a}_{2} \right)$.
These configurations are defined in Ref. \cite{Fan2014p023812} as: (i) a normal phase, where all order parameters are zero,$\langle \hat{\sigma}_{z} \rangle = \langle \hat{a}_{j}^{\dagger} \hat{a}_{j} \rangle = \langle \hat{\chi} \rangle = 0$, and corresponds to a separable ground state with zero excitation, (ii) two single-mode superradiant phases, where the qubit energy difference and one of the mean photon numbers are different from zero, $\langle \hat{\sigma}_{z} \rangle \neq 0$ with $ \langle \hat{a}_{1}^{\dagger} \hat{a}_{1} \rangle = \langle \hat{\chi} \rangle = 0$ and $\langle \hat{a}_{2}^{\dagger} \hat{a}_{2} \rangle \neq 0$ or  $\langle \hat{a}_{2}^{\dagger} \hat{a}_{2} \rangle = \langle \hat{\chi} \rangle = 0$ and $\langle \hat{a}_{1}^{\dagger} \hat{a}_{1} \rangle \neq 0$, (iii) and a two-mode superradiant phase, where all the order parameters are nonzero, $\langle \hat{\sigma}_{z} \rangle \neq 0 $,  $\langle \hat{a}_{j}^{\dagger} \hat{a}_{j} \rangle \neq 0$, and $\langle \hat{\chi} \rangle \neq 0$.
For example, in Ref. \cite{Fan2014p023812}, an exceptional solution is found under resonant fields, $\omega_{1}=\omega_{2}= \omega$, the two-mode superradiant phase appears just for equal couplings above a critical coupling, $g_{1}=g_{2}> g_{c} = \sqrt{\omega_{0}\omega}/2$.
Here, we are going to discard the two-mode photon number operator as order parameter and use the more adequate two-mode $SU(2)$ operator $\hat{J}_{x} = (\hat{a}^{\dagger}_{1} \hat{a}_{2} + \hat{a}_{1} \hat{a}^{\dagger}_{2} )/2$ that describes hopping between modes as we already know that the systems show such a partial symmetry.

\begin{figure}
\centering \includegraphics[scale= 1]{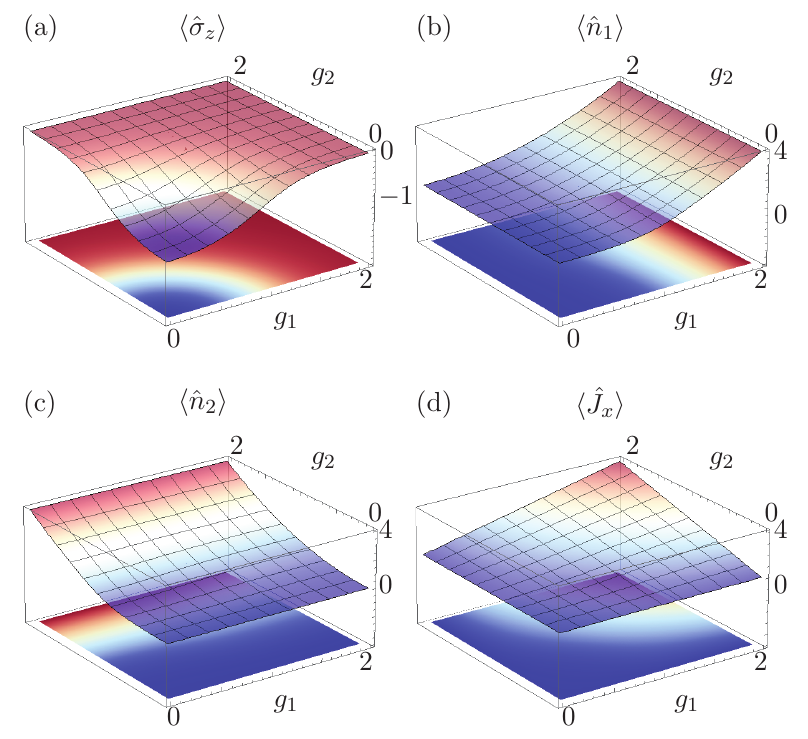}
\caption{(Color online) Four order parameters (a) mean energy difference,  $\langle \hat{\sigma}_{z} \rangle$, (b) first field mean photon number, $\langle \hat{a}^{\dagger}_{1} \hat{a}_{1} \rangle$, (c) second field mean photon number, $\langle \hat{a}^{\dagger}_{2} \hat{a}_{2} \rangle$ and (d) two-mode $SU(2)$ mean hopping,  $\langle \hat{J}_{x} \rangle$, for the ground state of Hamiltonian $\hat{H}_{1}$ on resonance, $\omega_{0} = \omega_{j} = \omega$. The couplings are given in units of the field frequency, $\omega$.} \label{fig:Figure1}
\end{figure}

\begin{figure}
\centering \includegraphics[scale= 1]{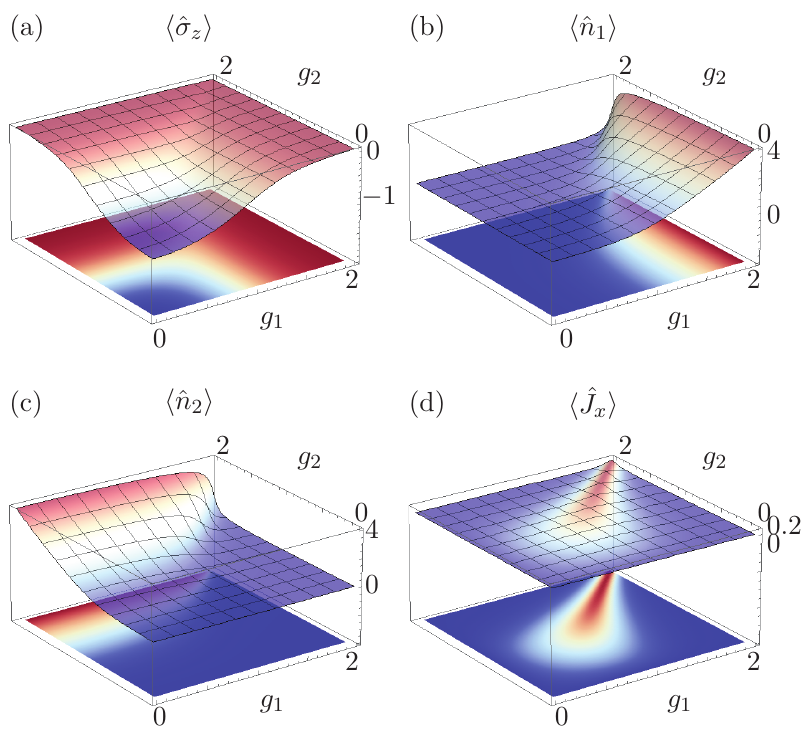}
\caption{(Color online) Same as Fig. \ref{fig:Figure1} for the ground state of $\hat{H}_{2}$ on resonance, $\omega_{0} = \omega_{j} = \omega$. The couplings are given in units of the field frequency, $\omega$.} \label{fig:Figure2}
\end{figure}

In order to find the ground state configuration for the single qubit models presented here, we set to study the exceptional case of resonant fields, $\omega_{1}= \omega_{2}$, that simplifies the problem by eliminating the field-field coupling,  
\begin{eqnarray}
\hat{H}_{1RF} &=& \hat{H}_{1D} \vert _{\omega_{1}= \omega_{2}}, \\
&=& \frac{\omega_{0}}{2} \hat{\sigma}_{z} + \omega \sum_{j=1}^{2} \hat{a}_{j}^{\dagger} \hat{a}_{j} + g \left( \hat{a}_{1}^{\dagger} + \hat{a}_{1} \right) \hat{\sigma}_{x}, \\
&=& \omega \hat{a}_{2}^{\dagger} \hat{a}_{2}  + \hat{H}_{R}
\end{eqnarray}
At this point, it is straightforward to recover the effective coupling transition for the quantum Rabi model in the weak coupling regime \cite{Buzek2005p163601,Tsyplyatyev2009p012134}, 
\begin{eqnarray}
g_{1c} = \frac{1}{2} \sqrt{\omega_{0} \omega},
\end{eqnarray}
and conclude that the model presents a ground state configuration with no excitation, $\vert g \rangle_{q} \vert 0 \rangle_{1} \vert 0 \rangle_{2}$, for coupling parameters in the range $0 \le \sqrt{g_{1}^{2} + g_{2}^{2}} < g_{1c}$.
This can be seen more clearly in Fig. \ref{fig:Figure1}(a), where the mean qubit energy levels difference is shown for $\hat{H}_{1RF}$ calculated with standard numerical methods \cite{RodriguezLara2010p2443,Robles2015p033819}.
Figure \ref{fig:Figure1}(b) and Fig. \ref{fig:Figure1}(c) show the mean photon number in the first and second field, in that order, and Fig. \ref{fig:Figure1}(d) shows the mean of the two-mode $SU(2)$ hopping operator.
It is also straightforward to borrow the deep-strong coupling, $g \gg \omega_{0}$, result from the literature \cite{Braak2011p100401,Braak2013p23} and realize that the ground state will be two-fold degenerate with ground state energy proportional to $- g^{2}/ \omega$, corresponding to the two parity separable ground states, $\frac{1}{\sqrt{2}} \left( \mp \vert e \rangle + \vert g \rangle \right) \left\vert \pm  \beta_{1} \right\rangle_{1} \left\vert \pm  \beta_{2} \right\rangle_{2}$ with the fields in coherent states, $\vert \beta \rangle = \sum_{n} (\beta^{n} / \sqrt{n!})\vert n \rangle$, with parameters $\beta_{j} = g_{j}/g$. 
This ground state configuration leads to mean values $\langle \hat{\sigma}_{z} \rangle = 0$, $\langle \hat{n}_{1} \rangle = \vert \beta_{1} \vert^2$, $\langle \hat{n}_{2} \rangle = \vert \beta_{2} \vert^2$ and $\langle \hat{J}_{x} \rangle =   2 \mathrm{Re}(\beta_{1}^{\ast}\beta_{2})$.
Note that, in the case of Hamiltonian $\hat{H}_{1}$, we can just describe two types of ground state configurations, the vacuum configuration where all four order parameters are zero, and a non-vacuum configuration where all four order parameters are different from zero.

In order to find the critical coupling for the second model for resonant fields,
\begin{eqnarray}
 \hat{H}_{2RF} & = &  \frac{\omega_{0}}{2} \hat{\sigma}_{z} + \omega \left( \hat{a}_{1}^{\dagger} \hat{a}_{1}  + \hat{a}_{2}^{\dagger} \hat{a}_{2} \right)  +  \nonumber \\ 
&& + \left[ g \hat{a}_{1}^{\dagger} + \frac{g_{1}^{2}- g_{2}^{2}}{g} \hat{a}_{1} - \frac{2 g_{1} g_{2}}{g} \hat{a}_{2} \right] \hat{\sigma}_{+} + \left[ \frac{g_{1}^{2}- g_{2}^{2}}{g} \hat{a}_{1}^{\dagger} +  g \hat{a}_{1} - \frac{2 g_{1} g_{2}}{g} \hat{a}_{2}^{\dagger} \right] \hat{\sigma}_{-} ,
\end{eqnarray} 
it is simpler to work with $\hat{H}_{2}$ in the zero and single excitation subspaces, then just half the critical coupling found \cite{Carmichael1973p47} to obtain,
\begin{eqnarray}
g_{2c} = g_{1c} .
\end{eqnarray}
Thus, the vacuum ground configuration will be in the parameter range $0 \le \sqrt{g_{1}^{2} + g_{2}^{2}} < g_{2c}$ and we recover the result in Ref. \cite{Fan2014p023812} for resonant fields and qubit.
Again, this is simpler to see in the mean qubit population inversion, Fig. \ref{fig:Figure2}(a).
Figure \ref{fig:Figure2} shows the four order parameters defined above for Hamiltonian $\hat{H}_{2}$.
Note that we recover the four ground state configurations described in Ref. \cite{Fan2014p023812} if we just exchange their two-mode photon number, $\hat{\chi}$, for the two-mode $SU(2)$ hopping operator, $\hat{J}_{x}$.
Thus, we are still able to see a change in the ground state configuration for just the single qubit in the case of identical couplings, $g_{1} = g_{2} > g_{2c}$, as expected from the semi-classical model analysis \cite{Fan2014p023812}.
We can write some of the different ground states for Hamiltonian $\hat{H}_{2}$. 
In the cases $g_{2} \ll g_{c}$ and $g_{1} > g_{c}$ or $g_{1} \ll g_{c}$ and $g_{2} > g_{c}$ the ground states will be given by $\frac{1}{\sqrt{2}} \left( \mp \vert e \rangle + \vert g \rangle \right) \left\vert \pm  g_{1} \right\rangle_{1} \left\vert   0 \right\rangle_{2}$  or $\frac{1}{\sqrt{2}} \left( \mp \vert e \rangle + \vert g \rangle \right) \left\vert  0 \right\rangle_{1} \left\vert  \pm g_{2} \right\rangle_{2}$ where the field states $\vert \pm g_{j} \rangle_{j}$ are coherent states.
In the regions $g_{1} \gg 2 g_{2}$ with $g_{2} \gg 0$ and $g_{2} \gg 2 g_{1}$ with $g_{1} \gg 0$ the ground state configuration will be given by that of Hamiltonian $\hat{H}_{1}$ in the deep strong coupling regime, $\frac{1}{\sqrt{2}} \left( \mp \vert e \rangle + \vert g \rangle \right) \left\vert \pm  \beta_{1} \right\rangle_{1} \left\vert \pm  \beta_{2} \right\rangle_{2}$ with the coherent parameters as defined beforehand.

Note that, as expected, in these single qubit models the transition between ground state configurations is smooth and can be understood as a quantum precursor of the phase transitions observed in the classical limit with an infinitely large ensemble of qubits.

\section{Spectra.}

\begin{figure}
\centering \includegraphics[scale= 1]{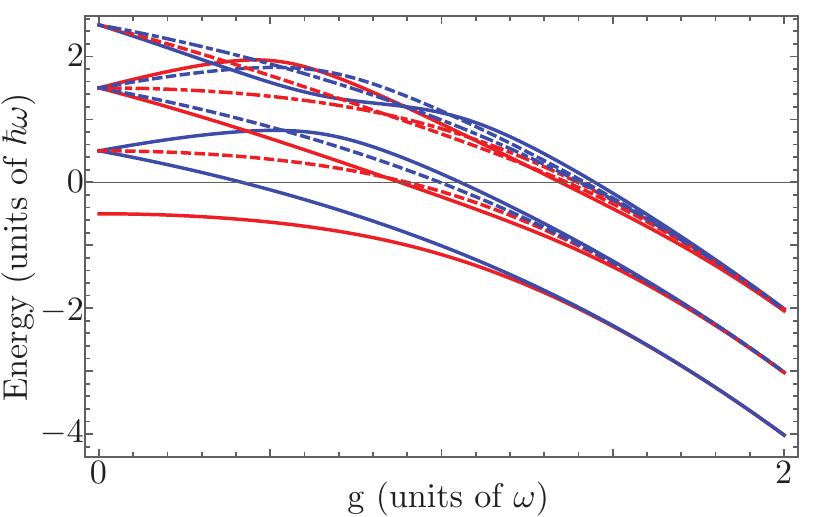}
\caption{(Color online) First twelve members of the spectra of Hamiltonian $\hat{H}_{1}$ with branches labeled as follow: Positive parity branches, $\left\{ \vert n_{b} \rangle_{2} \vert +, j \rangle \right\}$ are shown in red, negative parity branches, $\left\{ \vert n_{b} \rangle_{2} \vert -, j \rangle \right\}$, are shown in blue. The branches corresponding to $n_{b}=0,1,2$ are shown as solid, dashed and dot-dashed lines, in that order. } \label{fig:Figure3}
\end{figure}

Here, we will use continue our analysis of the exceptional case of resonant fields, $\omega_{1} = \omega_{2}= \omega$, to find a solution for the spectra of the models.
In the case of Hamiltonian $\hat{H}_{1}$, it is straightforward to see that the model is tractable in the exceptional case of resonant fields as in the displaced frame it can be written as $\hat{H}_{1RF}$. 
The proper basis for this Hamiltonian in the displaced frame is given by $\left\{ \vert n_{b} \rangle_{2} \vert \pm, j \rangle_{q,1} \right\}$, thus, each an every eigenstate can be labeled by the displaced mean photon number of the second field, $n_{b} = 0, 1, 2, \ldots$, and parity component, $(\pm, j)$ with $j=0,1, 2, \ldots$, of the Rabi basis \cite{Braak2011p100401,Braak2013p23,Braak2013p175301,Zhong2013p415302,Moroz2013p319,Maciejewski2014p16}.
Figure \ref{fig:Figure3} shows the first ten proper values of Hamiltonian $\hat{H}_{1}$, note that each and everyone of them can be labeled at any given effective coupling strength and the crossings in the spectra are always between different subspaces. 
The eigenvalues shown correspond to the following states, $\left\{ \vert 0 \rangle_{2} \vert +, j \rangle \right\}$ with $j=0,1,2$ in solid red lines, $\left\{ \vert 0 \rangle_{2} \vert -, j \rangle \right\}$ with $j=0,1,2$ in solid blue lines, $\left\{ \vert 1 \rangle_{2} \vert +, j \rangle \right\}$ with $j=0,1$ in dashed red lines, $\left\{ \vert 1 \rangle_{2} \vert -, j \rangle \right\}$ with $j=0,1$ in dashed blue lines, $\left\{ \vert 2 \rangle_{2} \vert +, 0 \rangle \right\}$ in dot-dashed red lines, and $\left\{ \vert 2 \rangle_{2} \vert -, 0 \rangle \right\}$ in dot-dashed blue lines.
As expected for large values of the effective coupling constant, $g \gg \omega$, the ground state will be twofold degenerate and the degeneracy of the rest will increase in twofold steps in the exceptional case of resonant fields; e.g., in this regime the two-fold degenerate ground state corresponds to $\left\{ \vert 0 \rangle_{2} \vert \pm, 0 \rangle  \right\}$, the four-fold first excited state, $\left\{ \vert 0 \rangle_{2} \vert \pm, 1 \rangle, \vert 1 \rangle_{2} \vert \pm, 0 \rangle  \right\}$, the six-fold second excited state, $\left\{ \vert 0 \rangle_{2} \vert \pm, 2 \rangle, \vert 1 \rangle_{2} \vert \pm, 1 \rangle, \vert 2 \rangle_{2} \vert \pm, 0 \rangle  \right\}$, and so on.

\begin{figure}
\centering \includegraphics[scale= 1]{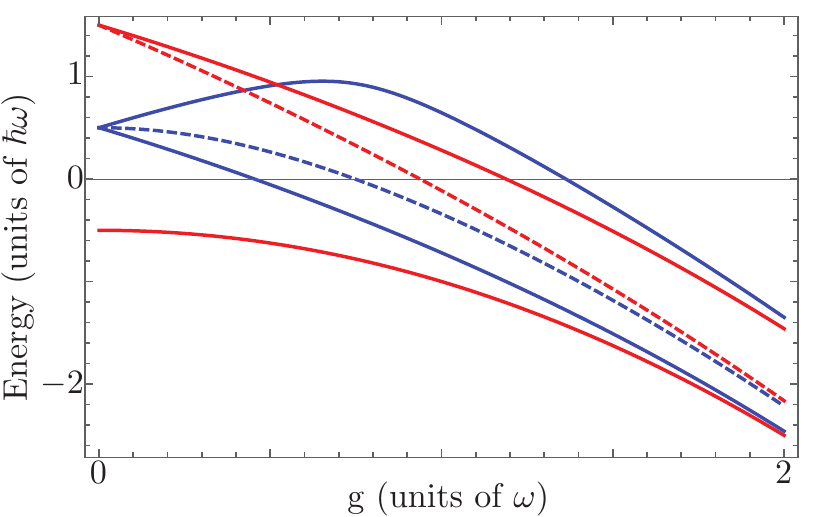}
\caption{(Color online) Six members of the spectra of Hamiltonian $\hat{H}_{1}$ with branches labeled as follow: The eigenvalues corresponding to $\vert + ,0, j \rangle$ with $j=0,1$ and $\vert + ,2, 0\rangle$ are shown in solid and dashed red, respectively, those related to $\vert - ,1, j\rangle$ with $j=0,1$ and $\vert - ,-1, 0\rangle$ in solid and dashed blue, in that order.  } \label{fig:Figure4}
\end{figure}

As we shown before, Hamiltonian $\hat{H}_{2}$ has four regimes, $\{ g_{1} \ll \omega, g_{2} \ll \omega,  g_{1} \gg 2 g_{2}, g_{2} \gg 2 g_{1} \}$, where it can be approximated by Hamiltonian $\hat{H}_{1}$. 
Thus, the spectra in these regimes can be labeled by the displaced second field photon number and Rabi basis as we have just shown above but there is a fifth regime $g_{1} = g_{2}$ where the spectra can be constructed. 
As mentioned in Sec. 2, the Hamiltonian $\hat{H}_{2}$ for the exceptional case of resonant fields and equal couplings conserves the  excitation operator $\hat{\mathcal{N}} = -\hat{a}_{1}^{\dagger} \hat{a}_{1} + \hat{a}_{2}^{\dagger} \hat{a}_{2} + \hat{\sigma}_{z}/2 + 1$.
This partitions the whole Hilbert space in subspaces of infinite dimension with the same mean value $\langle \hat{\mathcal{N}} \rangle = 0, \pm 1, \pm 2, \ldots$.
Note that we have the three labels we need to uniquely identify each eigenstate, $\left\{ \vert \pm, n_{d}, j \rangle \right\}$, the total parity of the state, $\langle \hat{\pi} \rangle = \pm$, the displaced excitation operator defining the subspace, $n_{d}=\langle \hat{\mathcal{N}} \rangle$, and the ordering in the $n_{d}$ subspace, $j$.
Figure \ref{fig:Figure4} shows six members of the spectra following this convention, the ground state will always be $\vert +, 0, 0 \rangle$ and will be completely degenerate at large effective coupling values $g_{1} \gg \omega$ due to the fact that each subspace has an infinite dimension.
In Fig. \ref{fig:Figure4} we also show the eigevalues corresponding to $\vert + ,0, j \rangle$ with $j=0,1$ and $\vert + ,2, 0\rangle$ in solid and dashed red lines, respectively, as well as those related to $\vert - ,1, j\rangle$ with $j=0,1$ and $\vert - ,-1, 0\rangle$ in solid and dashed blue lines, in that order.

\section{Single excitation dynamics}

\begin{figure}
\centering \includegraphics[scale= 1]{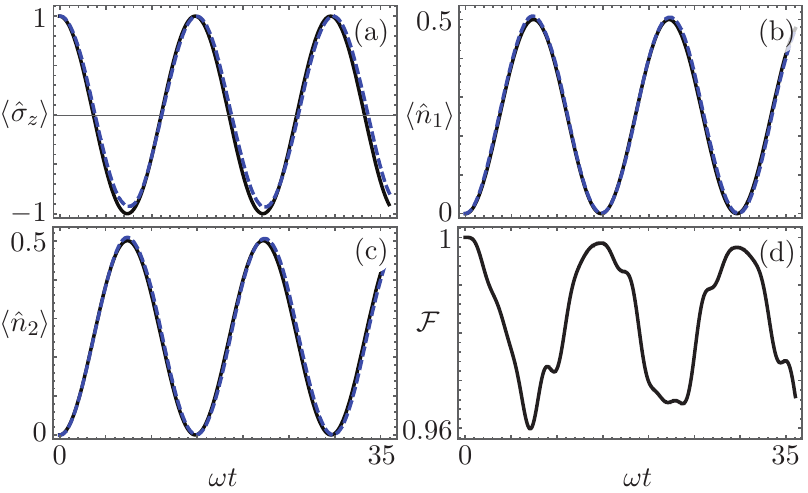}
\caption{(Color online) Time evolution of the (a) mean energy difference,  $\langle \hat{\sigma}_{z} \rangle$, (b) first field mean photon number, $\langle \hat{a}^{\dagger}_{1} \hat{a}_{1} \rangle$, (c) second field mean photon number, $\langle \hat{a}^{\dagger}_{2} \hat{a}_{2} \rangle$ and (d) fidelity,  $\mathcal{F}= \vert \langle \tilde{\psi}(t) \vert e^{- i \hat{H}_{1} t} \vert \psi(0) \rangle \vert^{2}$, for an initial state $\vert \psi (0) \rangle = \vert e \rangle_{q} \vert 0 \rangle_{1} \vert 0 \rangle_{2}$ under dynamics given by Hamiltonian $\hat{H}_{1}$ on resonance, $\omega_{0} = \omega_{j} = \omega$ and equal qubit-field couplings in the ultra-strong coupling regime, $g_{1}=g_{2}= 0.15 \omega$ that realizes a 50/50 beam splitter under weak coupling.} \label{fig:Figure5}
\end{figure}

\begin{figure}
\centering \includegraphics[scale= 1]{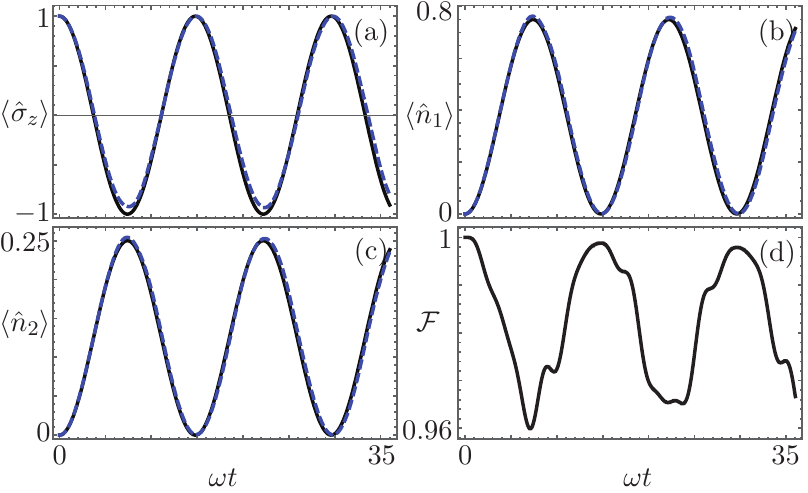}
\caption{(Color online) Time evolution of the (a) mean energy difference,  $\langle \hat{\sigma}_{z} \rangle$, (b) first field mean photon number, $\langle \hat{a}^{\dagger}_{1} \hat{a}_{1} \rangle$, (c) second field mean photon number, $\langle \hat{a}^{\dagger}_{2} \hat{a}_{2} \rangle$ and (d) fidelity,  $\mathcal{F}= \vert \langle \tilde{\psi}(t) \vert e^{- i \hat{H}_{1} t} \vert \psi(0) \rangle \vert^{2}$, for an initial state $\vert \psi (0) \rangle = \vert e \rangle_{q} \vert 0 \rangle_{1} \vert 0 \rangle_{2}$ under dynamics given by Hamiltonian $\hat{H}_{1}$ on resonance, $\omega_{0} = \omega_{j} = \omega$ and equal qubit-field couplings in the ultra-strong coupling regime, $g_{1}^{2}= 3 g_{2}^{2}$ with identical effective coupling to that in Fig. \ref{fig:Figure5}, $g^{2}= 0.212 \omega^{2}$, that realizes a 75/25 beam splitter under weak coupling.} \label{fig:Figure6}
\end{figure}

Dynamics of the quantum Rabi model are well studied in the most relevant regimes; weak \cite{Jaynes1963p89,Eberly1980p1323}, $g \ll w$, ultra-strong \cite{Ballester2012p021007}, $ g \gtrsim 0.1 \omega$, and deep-strong coupling \cite{Casanova2010p263603} , $g \ge \omega$, regimes \cite{Casanova2010p263603}.
Here, as a practical example, we consider the exceptional case of resonant fields in order to entangle the first and second cavities with a single excitation near the deep-strong coupling regime.
In the weak-coupling regime,  $g \ll w$, the time evolution operator for the resonant quantum-Rabi model is well known.
In the single excitation subspace, $\left\{ \vert e \rangle_{q} \vert 0 \rangle_{1}, \vert g \rangle_{q} \vert 1 \rangle_{1}  \right\}$, it yields the following time evolution for an initial state $\vert \psi (0) \rangle = \vert e \rangle_{q} \vert 0 \rangle_{1} \vert 0 \rangle_{2}$,
\begin{eqnarray}
\vert \tilde{\psi}_{1}(t) \rangle \approx \cos gt  \vert e \rangle_{q} \vert 0 \rangle_{1} \vert 0 \rangle_{2}  - i \sin gt \vert g \rangle_{q} \left[ \cos \xi \vert 1 \rangle_{1} \vert 0 \rangle_{2} + \sin \xi \vert 0 \rangle_{1} \vert 1 \rangle_{2} \right], \quad g \ll \omega.
\end{eqnarray}
The time evolved state oscillates between the original state and an entangled state of the two fields.
Thus, for a single excitation in the weak coupling regime, the first model can act as a $SU(2)$ beam splitter at times $t = n \pi/(2g)$. 
There, the qubit will be at the ground state and the cavities will be sharing a photon in a ratio $ \vert \cos \xi \vert^{2} / \vert \sin \xi \vert^{2} =  g_{1}^{2}/g_{2}^{2}$ given by the ratio between the two qubit-field couplings as $\tan \xi = g_{2}/g_{1}$.
Now, in the weak coupling regime, this process is slow, in order to obtain a fast beam splitter we need to go for stronger couplings. 
Figure \ref{fig:Figure5} shows the time evolution of the mean qubit population inversion and mean photon numbers of the fields for a 50/50 beam splitter realization in the ultra-strong regime, $g_{1}=g_{2}=0.15 \omega$ that gives $g^{2}=0.212 ~\omega^{2}$. 
The numerical evolution is compared to the result obtained in the weak coupling regime and it is possible to see that they are in close agreement during the first oscillation. 
If we stay with this effective coupling regime, $g^{2}= 0.212~ \omega^{2}$, we can keep this level of agreement for different splitting parameters, Fig. \ref{fig:Figure6} presents the case $g_{1}^{2} = 3 g_{2}^{2}$ that in the weak coupling regime realizes a 75/25 beam splitter.
In both cases we can see that the qubit does not reach complete transfer to the ground state, still a conditional measurement of the qubit in the ground state delivers a state close enough to the ideal split single photon state; Figures \ref{fig:Figure5}(d) and  \ref{fig:Figure6}(d) show the fidelity between the weak coupling evolution and the exact numerical ultra-strong coupling evolution, $\mathcal{F}=\vert \langle \tilde{\psi}_{1}(t) \vert e^{- i \hat{H}_{1}  t} \vert \psi(0) \rangle \vert^{2}$.

\begin{figure}
\centering \includegraphics[scale= 1]{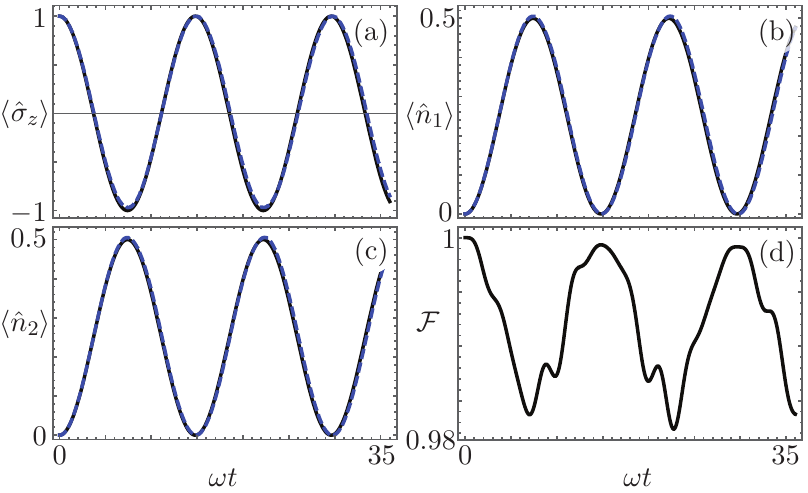}
\caption{(Color online) Same as Fig. \ref{fig:Figure5} under dynamics given by Hamiltonian $\hat{H}_{2}$.} \label{fig:Figure7}
\end{figure}

\begin{figure}
\centering \includegraphics[scale= 1]{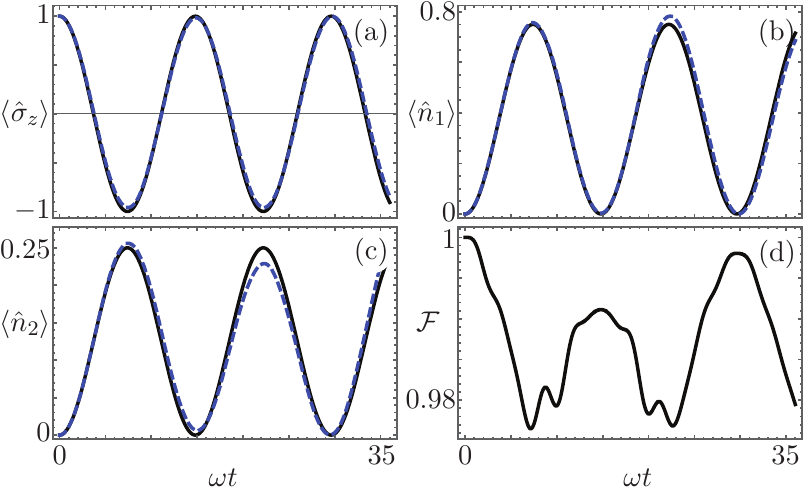}
\caption{(Color online) Same as Fig. \ref{fig:Figure6} under dynamics given by Hamiltonian $\hat{H}_{2}$.} \label{fig:Figure8}
\end{figure}

We can do the equivalent with Hamiltonian $\hat{H}_{2}$ \cite{GarciaMelgarejo2013,GarciaMelgarejo2013p31}, but in this case the one-excitation subspace for weak couplings, $g_{j} \ll \omega$, will be defined by the tripartite basis $ \{ \vert g \rangle_{q} \vert 1 \rangle_{1} \vert 0 \rangle_{2}, \vert e \rangle_{q} \vert 0 \rangle_{1} \vert 0 \rangle_{2}, \vert g \rangle_{q} \vert 0 \rangle_{1} \vert 1 \rangle_{2} \}$, and obtain the evolution for an initial state $\vert \tilde{\psi}(0) \rangle = \vert e \rangle_{q} \vert 0 \rangle_{1} \vert 0 \rangle_{2}$, up to an overall phase constant,
\begin{eqnarray}
\vert \tilde{\psi}_{2}(t) \rangle \approx \cos gt  \vert e \rangle_{q} \vert 0 \rangle_{1} \vert 0 \rangle_{2}  - i \sin gt \vert g \rangle_{q} \left[ \cos \xi \vert 1 \rangle_{1} \vert 0 \rangle_{2} - \sin \xi \vert 0 \rangle_{1} \vert 1 \rangle_{2} \right], \quad g \ll \omega.
\end{eqnarray}
Again, in the weak coupling limit, the time evolution under $\hat{H}_{2}$ dynamics delivers a single-photon beam splitter state of the two fields at the time  $t = n \pi/(2g)$. 
Figure \ref{fig:Figure7} and Fig. \ref{fig:Figure8} show the numerical exact evolution of the initial state in the ultra-strong coupling regime compared to the approximate result obtained using the weak coupling evolution for a 50/50 and 75/25 realization with the same parameters as those in Fig. \ref{fig:Figure5} and Fig. \ref{fig:Figure6}, in that order. 
Note that the fidelity in these cases is better than those under Hamiltonian $\hat{H}_{1}$ dynamics for short evolution times but it seems to degrade faster for longer evolution times.

\section{Conclusions}

We have studied exceptional solutions for two models describing a single two-level system coupled to two boson field modes. 
The first is a finite dissipative two-state system with only two fields in the absence of tunneling.
The second is equivalent to two-orthogonal fields coupling to the two corresponding orthogonal dipoles of a two-level system.
Both models may be feasible of experimental realization in cavity- or circuit-QED \cite{Fan2014p023812,Baust2014}.

The models conserve parity and show a partial SU(2) symmetry involving the two boson modes, thus, we explored regimes where they may be related to well known models with similar structure, like the quantum Rabi model.
We focused on the exceptional case of resonant fields where the models are analytically tractable.
Although only one of the models can be transformed to a form including the quantum Rabi model, we found that the ground state configurations of both models present the same critical coupling than the quantum Rabi model. 
Around this critical coupling, the ground state goes from the so-called normal configuration with no excitation, the qubit in the ground state and the fields in the quantum vacuum state, to a ground state with excitations, the qubit in a superposition of ground and excited state while the fields are not in the vacuum anymore, for the first model.
The second model shows a more complex ground state configuration landscape where we find the normal configuration just mentioned before, two single-mode configurations, where just one of the fields and the qubit are excited, and a dual-mode configuration, where both fields and the qubit are excited.
For the first model and some regions of the second model, we showed that the field components of the ground state are given by coherent states  in the deep-strong coupling regime. 
Following the integrability criteria for the quantum Rabi model established by Braak \cite{Braak2011p100401}, our results point to the integrability of these exceptional solutions in these regions.

We have also shown that these models for resonant fields and qubit frequencies can be used as two-mode $SU(2)$ beam splitters in the case of single excitation even in the ultra-strong coupling regime; at specific times, an excited two-level system may give its energy quanta to the two commuting boson fields, this excitation will be shared in an entangled state by the two fields in a manner dictated by their coupling ratio.



\end{document}